\documentclass{article}

\usepackage{amsmath,amssymb,amsthm}
\usepackage{graphicx}
\usepackage{caption}
\usepackage{subcaption}
\usepackage{url}
\usepackage{a4wide}

\theoremstyle{definition}

\begin{document}

\title{Application of SIR epidemiological model: new
trends\thanks{This is a preprint of a paper whose
final and definite form is in International Journal of Applied Mathematics and Informatics.
Please cite this paper as: \emph{Rodrigues, Helena Sofia (2016). Application of SIR epidemiological model: new
trends, International Journal of Applied Mathematics and Informatics, 10: 92--97.}} }
\author{Helena Sofia Rodrigues$^{1,2}$\\
{\tt \small sofiarodrigues@esce.ipvc.pt}
}

% ----------

\date{$^1$ \mbox Business School, Viana do Castelo Polytechnic Institute,\\
Portugal\\[0.3cm]
$^2${\text{Center for Research and Development in Mathematics and Applications (CIDMA)}},
University of Aveiro, Portugal\\[0.3cm]
}

\maketitle

% ----------

%-----------------------------------------------------------------------------------------

\begin{abstract}
The simplest epidemiologic model composed by mutually exclusive
compartments SIR (susceptible-infected-susceptible) is presented
to describe a reality. From health concerns to situations related with
marketing, informatics or even sociology, several are the
fields that are using this epidemiological model as a first
approach to better understand a situation. In this paper, the basic transmission
model is analyzed, as well as simple tools that allows us to extract
a great deal of information about possible solutions. A set of applications -
traditional and new ones - is described to show the importance of this model.

\smallskip


\end{abstract}

%
%  Main text of the article.
%

\section{Introduction}
Epidemiology has become an important issue for modern society. The relationship between mathematics
and epidemiology has been increasing. For the mathematician, epidemiology provides new and exciting
branches, while for the epidemiologist, mathematical modeling offers an important research tool in the
study of the evolution of diseases.

The SIR model, developed by Ronald Ross, William Hamer, and others in the early twentieth century
\cite{Anderson1991}, consists of a system of three coupled nonlinear ordinary differential equations.

Theoretical papers by Kermack and McKendrinck, between 1927 and 1933
about infectious disease models, have had a great influence in the deve\-lopment
of mathematical epidemiology models \cite{Murray2002}. Most of the basic theory
had been developed during that time, but the theoretical progress has been steady since then
\cite{Brauer2008}. Mathematical models are being increasingly used
to elucidate the transmission of several diseases. These models,
usually based on compartment models, may be rather simple,
but studying them is crucial in gaining important knowledge of the
underlying aspects of the infectious diseases spread out \cite{Hethcote1994},
and to evaluate the potential impact of control programs
in reducing morbidity and mortality.

Recent years have seen an increasing trend in the representation
of mathematical models in publications in the epidemiological
literature, from specialist journals of medicine, biology and
mathematics to the highest impact generalist journals
\cite{Ferguson2006}, showing the importance of interdisciplinary.

But this epidemiological model crossed the borders of health and biology. In several fields,
the concept of spreading is applied and,  a pragmatic point of view, the SIR model
is a beginning point to understand what happens rapidly; then, with more understanding and
complexity is possible to enrich the model and put more details in the formulation.

The paper is organized as follows. Next section the SIR model is presented,
as well as some theoretical results that allows to understand the transmission process.
Then a set of application from distinct fields are exposed in Section 3 and, finally,
some concluding remarks are done.

\section{SIR model}

Mathematical models are a simplified representation of how an infection
spreads across a po\-pulation over time.

Most epidemic models are based on dividing the population
into a small number of compartments. Each containing individuals
that are identical in terms of their status with respect
to the disease in question. In the SIR model, the three compartments are:

\begin{itemize}
\item \emph{Susceptible} ($S$): is the class of individuals who are susceptible to infection;
this can include the passively immune once they lose their immunity or,
more commonly, any newborn infant whose mother has never been infected
and therefore has not passed on any immunity;

\item \emph{Infected} ($I$): in this class, the level of parasite is sufficiently
large within the host and there is potential in transmitting
the infection to other susceptible individuals;

\item \emph{Recovered or Resistant} ($R$):
includes all individuals who have been infected and have recovered.
\end{itemize}

This epidemiological model captures the dynamics of acute infections
that confers lifelong immunity once recovered. Diseases where individuals
acquire permanent immunity, and for which this model may be applied,
include measles, smallpox, chickenpox, mumps, typhoid fever and diphtheria.

Generally, the total population size is considered constant, \emph{i.e.}, $N=S+I+R$.
Then two cases should be studied, distinguished by the inclusion or exclusion of demographic factors.

\subsection{The SIR model without demography}

Having compartmentalized the population, we now need a set
of equations that specify how the sizes of compartments change over time.

\medskip

The SIR model, excluding births and deaths, can be defined as
\begin{equation}
\label{cap3:SIR_EDO_without_demography}
\begin{tabular}{l}
$\displaystyle\frac{dS}{dt}=-\beta S I$\\
\\
$\displaystyle\frac{dI}{dt}=\beta S I -\gamma I$\\
\\
$\displaystyle\frac{dR}{dt}=\gamma I$\\
\end{tabular}
\end{equation}
\noindent subject to initial conditions $S(0)>0$, $I(0)\geq0$ and $R(0)\geq0$.

In addition, the transmission rate, per capita, is $\beta$ and the recovery rate
is $\gamma$.

There are three commonly used threshold values in epidemiology:
$\mathcal{R}_0$, $\sigma$ and $R$. The most common and probably
the most important is the basic reproduction number
\cite{Heffernan2005,Hethcote2000,Hethcote2008}.
The basic reproduction number,
denoted by $\mathcal{R}_0$, is defined as the
average number of se\-condary infections that occurs when one
infective is introduced into a completely susceptible population.

This threshold, $\mathcal{R}_0$, is a famous result due
to Kermack and McKendrick \cite{Kermack1927} and is referred to
as the ``threshold phenomenon'', giving a borderline between
a persistence or a disease death.  $\mathcal{R}_0$ it is also
called the basic reproduction ratio or basic reproductive rate.

\medskip

The contact number, $\sigma$ is the average number
of adequate contacts of a typical infective during
the infectious period.

An adequate contact is one that is sufficient for transmission,
if the individual contacted by the susceptible is an infective.
It is implicitly assumed that the infected outsider is in the host population for
the entire infectious period and mixes with the host population in exactly
the same way that a population native would mix.

\medskip

The replacement number, $R$, is the average number of secondary
infections produced by a typical infective
during the entire period of infectiousness.

Note that the replacement number $R$ changes
as a function of time $t$ as the disease evolves after the initial invasion.

These three quantities $\mathcal{R}_0$, $\sigma$ and $R$
are all equal at the beginning of the spreading
of an infectious disease when the entire population (except the infective
invader) is susceptible. $\mathcal{R}_0$ is only defined at the time of invasion,
whereas $\sigma$ and $R$ are defined at all times.

The replacement number $R$ is the actual number of secondary cases from
a typical infective, so that after the infection has invaded a population and
everyone is no longer susceptible, $R$ is always less than the basic reproduction
number $\mathcal{R}_0$. Also after the invasion, the susceptible fraction is less
than one, and as such not all adequate contacts result in a new case. Thus the
replacement number $R$ is always less than the contact number $\sigma$ after the
invasion \cite{Hethcote2000}. Combining these results leads to
$$\mathcal{R}_0 \geq \sigma \geq R.$$

Note that $\mathcal{R}_0 = \sigma$ for most models,
and $\sigma > R$ after the invasion for all models.

For the models throughout this study the basic reproduction number,
$\mathcal{R}_0$, will be applied. When
$$\mathcal{R}_0 < 1$$
\noindent the disease cannot invade the population and the infection will die out over
a period of time. The amount of time this will take generally depends on how
small $\mathcal{R}_0$ is. When
$$\mathcal{R}_0 > 1$$
\noindent invasion is possible and infection can spread through the population. Generally,
the larger the value of $\mathcal{R}_0$ the more severe, and possibly widespread,
the epidemic will be \cite{Driessche2002}.

In this SIR model, when a newly introduced infected individual
can be expected to infect other people
at the rate $\beta$ during the expected infectious period $1/\gamma$. Thus,
this first infective individual can be expected to infect
$$\mathcal{R}_0=\frac{\beta}{\gamma}.$$

\subsection{The SIR model with demography}

The simplest and most common way of introducing demography into the SIR model
is to assume there is a natural host lifespan, $1/\mu$ years. Then, the rate
at which individuals, at any epidemiological compartment, suffer natural mortality
is given by $\mu$. It is important to emphasize that this factor is independent
of the disease and is not intended to reflect the pathogenicity of the infectious agent.
Historically, it has been assumed that $\mu$ also represents the population's crude birth rate,
thus ensuring that total population size does not change through time, or in other words,
$\frac{dS}{dt}+\frac{dI}{dt}+\frac{dR}{dt}=0$.

So, the SIR model, including births and deaths, can be defined as
\begin{equation}
\label{cap3:SIR_EDO_with_demography}
\begin{tabular}{l}
$\displaystyle\frac{dS}{dt}=\mu-\beta S I-\mu S$\\
\\
$\displaystyle\frac{dI}{dt}=\beta S I -\gamma I-\mu I$\\
\\
$\displaystyle\frac{dR}{dt}=\gamma I-\mu R$\\
\end{tabular}
\end{equation}
\noindent with initial conditions $S(0)>0$, $I(0)\geq 0$ and $R(0)\geq 0$.

It is important to introduce the $\mathcal{R}_0$ expression for this model.
The parameter $\beta$ represents the transmission rate per infective and the negative terms
in the equation tell us that each individual spends an average $\frac{1}{\gamma+\mu}$
time units in this class. Therefore, if we assume the entire population is susceptible,
then the average number of new infectious per infectious individual
is determined by
$$\mathcal{R}_0=\frac{\beta}{\gamma+\mu}.$$

The inclusion of demographic dynamics may allow a disease to die out or persist
in a population in the long term. For this reason it is important to explore
what happens when the system is at equilibrium.

A model defined SIR has an equilibrium point,
if a triple $E^{*}=\left(S^{*},I^{*},R^{*}\right)$ satisfies the following system:
$$
\begin{cases}
\frac{dS}{dt}=0\\
\frac{dI}{dt}=0\\
\frac{dR}{dt}=0\\
\end{cases}.
$$

If the equilibrium point has the infectious component equal to zero ($I^{*}=0$),
this means that the pathogen suffered extinction and $E^{*}$
is called Disease Free equilibrium (DFE).

If $I^{*}>0$ the disease persist in the population
and $E^{*}$ is called Endemic Equilibrium (EE).

\bigskip

With some calculations and algebraic manipulations, it is possible
to obtain two equilibria for the system (\ref{cap3:SIR_EDO_with_demography}):
\begin{equation*}
\label{cap3:SIR_equilibrium}\index{Disease Free Equilibrium}\index{Endemic Equilibrium}
\begin{tabular}{ll}
DFE: & $E^{*}_{1}=(1,0,0)$ \\
EE: & $E^{*}_{2}=\left(\frac{1}{\mathcal{R}_0},\frac{\mu}{\beta}\left(\mathcal{R}_0-1\right),
1-\frac{1}{\mathcal{R}_0}-\frac{\mu}{\beta}\left(\mathcal{R}_0-1\right)\right)$ \\
\end{tabular}
\end{equation*}

When $\mathcal{R}_0<1$, each infected individual produces, on average,
less than one new infected individual, and therefore, predictable that the infection
will be cleared from the population. If $\mathcal{R}_0>1$, the pathogen
is able to invade the susceptible population \cite{Heffernan2005,Hethcote2000}.
It is possible to prove that for the Endemic Equilibrium\index{Endemic Equilibrium}
to be stable, $\mathcal{R}_0$ must be greater than one, otherwise the
Disease Free Equilibrium is stable. More detailed information about local
and global stability of the equilibrium point\index{Equilibrium point}
can be found in \cite{Chavez2002,Kamgang2008,Li1996,Muldowney1999}.

This threshold behavior is very useful, once we can determine which control measures,
and at what magnitude, would be most effective in reducing $\mathcal{R}_0$
below one, providing important guidance for public health initiatives.

Next section some applications of this epidemiological model are presented,
as well as a set of references that can complement the information given.

\section{Applications}

\subsection{Health}
\subsubsection{Influenza}
\begin{figure}[ptbh]
\center
  \includegraphics [scale=0.6]{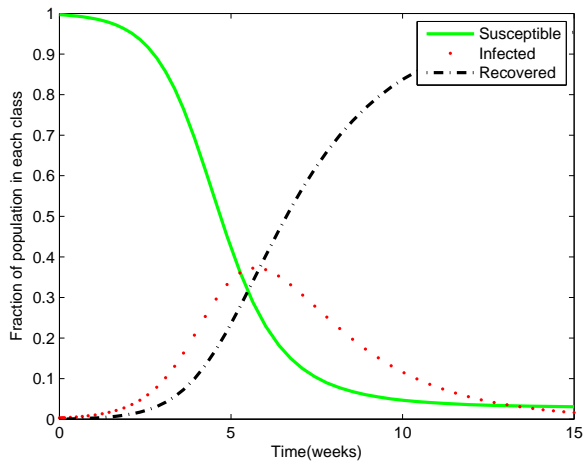}\\
   {\caption{\label{cap3_SIR_without_demography} The time-evolution of influenza over 15 days }}
\end{figure}

Consider an epidemic of influenza in a British boarding school \cite{Keeling2008}.
Three boys were reported to the school infirmary with the typical symptoms of influenza.
Over the next few days, a very large fraction of the 763 boys in the school had contact
with the infection. Within two weeks, the infection had become extinguished.
The best fit parameters yield an estimated active infectious period of $1/\gamma=2.2$
days and a mean transmission rate $\beta=1.66$ per day. Therefore, the estimated $\mathcal{R}_0$
is 3.652. Figure~\ref{cap3_SIR_without_demography} represents the dynamics of the three state variables.
It can be observed that the curve of susceptible is decreasing all over the time, because the birth
was no considered, and once become infected never returns to the state of susceptible.
The curve of infected reaches to a peak of the disease beyond 5 weeks. This information could be very
useful for health authorities to ensure that all resources are available - medicines,
doctors , hospitalization resources -  to provide a good
health care if necessary. Depending of flatness of the curve the response should be adaptive.
The curve related to the recovered compartment is important because
accumulates the number of individuals that have been seek in that outbreak.

More information about this disease and other authors that studied
influenza can be found in \cite{Hooten2010, Karim2011, Nichol2010}.

In other examples the curves do not be shown, instead the only thing that are
changed are the parameters values. The main goal of this paper is not to show all the graphics related to
SIR model, but to present a set of applications in several fields.

\medskip

\subsubsection{Dengue Fever}

Dengue is a vector-borne disease transmitted from an infected human to a female \emph{Aedes} mosquito by
a bite. Then, the mosquito, that needs regular meals of blood to feed their eggs, bites a potential healthy
human and transmits the disease making it a cycle. Nowadays, Dengue is the mosquito-borne infection
that has become a major international public health
concern. According to the World Health Organization (WHO), 50 to 100 million Dengue Fever infections
occur yearly, including 500000 Dengue Hemorrhagic Fever cases and 22000 deaths, mostly among children
\cite{WHO}.

This global pandemic is attributed to the unprecedented population
growth, the rising level of urbanization without adequate domestic water supplies, increasing movement of the virus
between humans (due to tourism, migration, or international trade), and lack of effective mosquito control. Dengue
virus is transmitted to humans through the bite of infected Aedes mosquitoes, specially Aedes Aegypti. Once infected,
a mosquito remains infected for life, transmitting the virus to susceptible individuals during feed. Without a vaccine,
vector control remains the only available strategy against dengue.
Appropriate mathematical models can give a deeper insight into the mechanism of disease transmission.

In this particular disease, the SIR model associated to the human population,
usually is coupled to a SI model for the mosquito, due to the vector transmission process.
More details, can be found in \cite{Sofia2009}-\cite{Sofia2015}.

\medskip

\subsubsection{SARS}

The Severe Acute Respiratory Syndrome (SARS) was the first epidemic of the 21st century.
It emerged in China late 2002 and quickly spread to 32 countries causing more than 774
deaths and 8098 infections worldwide \cite{Ng2003}.

SARS is a highly contagious respiratory
disease which is caused by the SARS Coronavirus. It is a serious form of pneumonia,
resulting in acute respiratory distress and sometimes death.
The SARS epidemic originated in China, in late 2002. Although the Chinese government
tried to control the the outbreak of the SARS epidemic without the awareness of the
World Health Organization (WHO), it continued to spread.

In the research papers \cite{Mkhatshwa2011} and \cite{Small2004} they use the SIR model, as a first
approach to explain this disease. The use the super-spreading individuals - infected individuals
that infect more than the average number of secondary cases - to modified the traditional
epidemiological model. The effect of super-spreaders can be used in cases where there is
a higher transmission rate.

\subsection{Networks}

\subsubsection{Online social networks}

The last decade has rise a huge number of online social networks (such as Facebook,
Twitter, MySpace, Instagram, Linkedin,...). Several papers have studied,
under epidemiological models, the adoption or abandonment of online social networks.
Cannarella and Spechler \cite{Cannarella2014} studied the information diffusion on Twitter, in order understand
the properties of underlying media and model communication patterns; with the popularity of Twitter
it become a venue to broadcast rumors and misinformation.

Wang and Wang \cite{Wang2015} investigate a SIR model to study rumor spreading.
With the development of microblogging technology, it become easy to publish several
messages on the network websites, and also for other people to be able to visit these websites
to search for messages according to their own needs, increasing rapidly the social network.

\medskip

\subsubsection{Viral Marketing}

Viral marketing (VM) is a recent approaching to markets and can potentially reach a large and fast
audience, through a cheap communication campaigns. VM exploits existing social networks by
encouraging people to share product information and campaigns with their
friends, through email or networks medium. This type of communication has more impact in the customer,
because the information was
recommended by friends and peer networks that knows the personal interests, instead of standard companies; this kind of communication have more impact because
is directly targeted. Besides
When a marketing message goes viral, it is analogous to an epidemic, since involves a person-to-person transmission, spreading
within a population. Rodrigues and Fonseca \cite{SofiaFonseca2016} explored a set of
simulation experiments to explore the influence of several controlled and
external factors that could influence viral campaigns.

Also known as internet worth of mouth marketing, VM has been gaining more fans, from professionals
to researchers, as an alternative strategy to traditional communication, transferring founds from
companies to online marketing actions and exploring this spreading phenomena
\cite{ Kandhaway2014, SofiaFonseca2015}.

\medskip

\subsubsection{Audience applause}

The social identity and crowd psychology study how and why an individual
change their behavior in response to others; within a group, a distinct attitude
can arise in a few persons and then spread quickly
to all other members.

According to Mann \emph{et al.}\cite{Mann2013} individuals' probability of starting
clapping increased in proportion to the
number of other audience members already affected by this social contagion.
In this paper, the authors apply a Bayesian model selection approach to
determine the dynamics of how some details or social cues can provoke the spread
of social behavior in a group of people. They reach to the conclusion that the
the audience clapping can vary, even when the quality of the presentations are identical,
changing according to the set of infected people.

\medskip

\subsubsection{Diffusion of ideas}

The population dynamics underlying the diffusion of ideas hold many
qualitative similarities to those involved in the spread of infections.
Bettencourt \emph{et al.} \cite{Bettencourt2006} explore this point of view
as a tool to quantify sociological and behavioral patterns. They explore the
spreading of Feynaman diagram through the theoretical physics communities of
the USA, Japan, and the USSR in the period immediately after World War II; having
this in mind they investigate the effectiveness of the adoption
of an idea, finding values for parameters that describe intentional social organization and long lifetimes for the idea.

By other hand, Funk \cite{Funk2013}, explore the concept of epidemiology in the human behavior
when public campaigns and mass media reports are diffused. The spread of awareness is crucial in this
model to describe the susceptible person to become convinced or informed to the disease and have additional
precautions related to the disease transmission process.

\subsection{Informatics}

\subsubsection{Peer-to-peer (P2P) newtworks}

Understanding the spread of information on complex networks is crucial from a
theoretical and applied perspective. To evaluate them with large-scale real-world data remains
an important challenge.

During the downloading process, the peer shares the downloaded parts of the
file and, thus, contributes to distributing it in the network \cite{Leibnitz2006}.
The authors consider a fie sharing application similar to eDonkey
which belongs to the class of hybrid P2P architectures and apply the SIR model, that  corresponds
to the populations of idle peers, peers currently downloading the file, and those sharing it.

Bernardes \emph{et al.} \cite{Bernardes2012} asses the relevance of the SIR model
to mimic key properties of spreading cascade of a file sharing.

\medskip

\subsubsection{Spread of computer virus}

Nowadays, with the rapid development of network information technology, information
networks security has become a very critical issue in our work and daily life. The computer
virus are being developed simultaneously with the computer systems and the use
of internet facilities increases the number of damaging virus incidents, producing serious
problems for individuals and corporate computer systems.
Antivirus software is the major means of defending against viruses. Although, antivirus
technique cannot predict the evolution trend of viruses and, hence, cannot provide
global suggestions for their prevention and control.
The strong desire to understand the spread mechanism of computer viruses has motivated
the proposal of a variety of epidemic models that are based on fully connected
networks, that is, networks where each computer is equally likely to be accessed by any
other computer.

Computer virus is considered as one of the most important weapon in the internet,
and their emergence and spread may have great effect on the computer world.
Different codes have different ways to spread
in the internet. Virus mainly attack the file system and worm uses system vulnerability
to search and attack the computer. And for trojan horses, they camouflage themselves and
thus induce the users to download them. There are a variety of computer virus, but they
all have infectivity, invisibility, latent, destructibility and unpredictability [7].
The word latent means that the virus hide themselves in the computer and spread them
in the internet while the users can not notice them. More details about applications to the spread of
computer virus can be found in \cite{Deng2015, Gan2012, Han2010, Zhu2012}.

\subsection{Economics and Finance}

\subsubsection{Rational expectations}

The economic epidemiology merges the epidemiological models with economic choice, translating a rational
decision making.
Economic research in this area began in response to the AIDS epidemic and has led to an improved
understanding of the thought/decisions towards a infectious disease, by anonymous individuals or policymakers
\cite{Aadland2011}. the power to eradicate an infectious disease is not only in the hands of policymakers or
health authorities: it is also important that rational individuals made their own response
to lower the prevalence of a disease, by increasing protection.
Economic epidemiology has made significant advances in educating health officials about the behavioral
implications of public policies. Aadland \emph{et al.} explored the nature of the short-run
equilibrium dynamics for rational expectations
economic epidemiological systems. They show
that well-intentioned policy has the potential to create instability when people behave
rationally and in a self-interested manner.

\subsubsection{Financial network contagion}

The financial sector is always a theme of interest, due to its importance in economy in general, and
our daily lives in particular.
Some papers \cite{Fisher2013, Toivanen2013} analyze the importance of individual
bank-specific factors on financial stability. The spreading of the contagion in
the interbank network can be seen as an epidemiological model.
The authors investigate the systemic risk and how this risk can propagate in
different bank and countries within the euro area.
Fisher makes counterfactual simulations to propagate shocks
emerging from three sources of systemic risk: interbank, asset price, and sovereign
credit risk markets. when the conditions deteriorate, these channels trigger severe direct
and indirect losses and cascades of defaults, whilst the dominance of the sovereign credit
risk channel amplifies, as the primary source of financial contagion in the banking network.

\subsection{Science Fiction: Zombies attack}
In 2009, the first mathematical investigation of the zombie community appears.
Taking their cues from traditional zombie movies, Munz \emph{et al.} \cite{Munz2009} hypothesized
the effect of a zombie attack and its impact on human civilization.
According to their mathematical model, ``a zombie outbreak is likely to lead to the collapse of civilization,
unless it is dealt with quickly. While aggressive quarantine
may contain the epidemic, or a cure may lead to coexistence
of humans and zombies, the most effective way to contain the rise of the undead is to hit hard and hit often.''
The model showed two equilibria: the disease-free equilibrium (with no zombies)
and the doomsday equilibrium (where everyone is a zombie).
The application of a linear stability analysis showed that -
in the absence of further interventions - the disease-free
equilibrium was unstable and the doomsday equilibrium was stable.
Since this paper, other authors follow this area with a careful attention (see more in \cite{Crossley2011, Langtangen2013}),
not only motivated by the tv series, but as a way to motivate young students for the epidemiology issues.

\section{Conclusion}

Deterministic models applied to the study of infectious disease have a long tradition.
The importance to predict the evolution of a disease, its impact in population and in the health systems
 - human and material resources - has a long concern to human population.
However, with the increasing of new technologies and the growth of the interdisciplinarity, new researchers
are become interested in epidemiological models to apply in other fields.

In this paper a selection of application from different fields was presented where the SIR models was used.
This exposition
is not exhaustive, but is a selection of recent areas that are been developed and
where the epidemiological mathematics is a possible response to describe reality.

Networks
and the epidemiology of directly transmitted infectious diseases are fundamentally
linked. Most of the themes that involve the diffusion phenomena could start
to simulate and understand some scenarios with
simple epidemiological models. They can be seen as a first tool to try when a new
problem presents itself, due to its limitations. But as simple as they seem, they are a
huge help to define new step in research and an emergency response to a crisis
when time to predict is short.

\section*{Acknowledgment}
This work was supported in part by the Portuguese Foundation for Science
and Technology (FCT-Fundação para a Ciência e a Tecnologia), through CIDMA
- Center for Research and Development in Mathematics and Applications,
within project UID/MAT/04106/2013.

\end{document}